\documentclass[
reprint,
onecolumn,
amsmath,amssymb,
superscriptaddress,aps,
pre
]
{revtex4}

\usepackage[utf8]{inputenc}
\usepackage{graphicx}
\usepackage{xcolor}
\usepackage{hyperref}
\usepackage{url}
\usepackage[normalem]{ulem}

\usepackage{amsmath}
\usepackage{amsfonts}
\usepackage{amssymb}
\usepackage{graphicx}

\begin{document}
\title{Collective dynamics of capacity-constrained ride-pooling fleets}

\author{Robin M. Zech}
\affiliation{Chair for Network Dynamics, Center for Advancing Electronics Dresden (cfaed) and Institute for Theoretical Physics, Technische Universit\"at Dresden, 01062 Dresden, Germany}

\author{Nora Molkenthin}
\affiliation{Complexity Science, Potsdam Institute for Climate Impact Research, Member of the Leibniz Association, 14473 Potsdam, Germany}

\author{Marc Timme}
\affiliation{Chair for Network Dynamics, Center for Advancing Electronics Dresden (cfaed) and Institute for Theoretical Physics, Technische Universit\"at Dresden, 01062 Dresden, Germany}

\author{Malte Schröder}
\affiliation{Chair for Network Dynamics, Center for Advancing Electronics Dresden (cfaed) and Institute for Theoretical Physics, Technische Universit\"at Dresden, 01062 Dresden, Germany}
\thanks{malte.schroeder@tu-dresden.de}

\begin{abstract}
Ride-pooling (or ride-sharing) services combine trips of multiple customers along similar routes into a single vehicle. 
The collective dynamics of the fleet of ride-pooling vehicles fundamentally underlies the efficiency of these services. In simplified models, the common features of these dynamics give rise to scaling laws of the efficiency that are valid across a wide range of street networks and demand settings. However, it is unclear how constraints of the vehicle fleet impact such scaling laws. 
Here, we map the collective dynamics of capacity-constrained ride-pooling fleets to services with unlimited passenger capacity and identify an effective fleet size of available vehicles as the relevant scaling parameter characterizing the dynamics. Exploiting this mapping, we generalize the scaling laws of ride-pooling efficiency to capacity-constrained fleets. We approximate the scaling function with a queueing theoretical analysis of the dynamics in a minimal model system, thereby enabling mean-field predictions of required fleet sizes in more complex settings. These results may help to transfer insights from existing ride-pooling services to new settings or service locations.
\end{abstract}

\maketitle

\section*{Introduction}
Human mobility is a quintessential example of a complex system \cite{holovatch2017complex, barbosa2018_mobility_models}. Interactions of individual travelers with each other, with their environment or with transportation services give rise to complex emergent mobility patterns and collective dynamics \cite{barbosa2018_mobility_models, helbing2000freezing, erhardt2019transportation, schroder2020anomalous, storch2021_Incentives}. Statistical physics approaches have helped to reveal universal patterns in the scaling of human mobility \cite{simini2012universal, barbosa2018_mobility_models}, characterize recurring aspects of the structure of mobility and transportation networks \cite{gastner2006optimal, verma2016emergence, barthelemy2008modeling, brelsford2018toward, xu2020deconstructing}, and explain fundamental properties of congestion and its persistence across a variety of systems \cite{helbing2000freezing, Karamouzas2014UniversalPedestrian, treiber2013_traffic, loder2019understanding, saberi2020simple, marszal2021phase}. Currently, human mobility is transforming towards new modes of transport that are increasingly self-organized and networked \cite{barbosa2018_mobility_models, mckinsey2019_mobility, erhardt2019transportation, schroder2020anomalous, storch2021_Incentives}. In particular, app-based on-demand ride-pooling services promise to reduce the economic and ecological impact of congestion and emissions in urban mobility, especially in light of the current trend of ongoing urbanization \cite{un2014_urbanizationProspect, un2018_urbanizationProspect_keyFacts, mcdonnell2016_urbanEcology, ramaswami2016_metaPrincipleSustainableCities}. 

\begin{figure}[!h]
    \centering
    \includegraphics{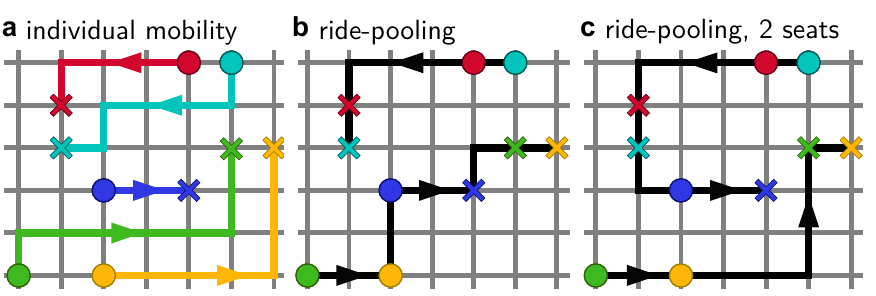}
    \caption{
    \textbf{Constraints shape the dynamics of ride-pooling.}
    \textbf{a} With individual mobility, each person travels from their origin (circle) to their destination (cross) using their own car (colored lines).
    \textbf{b} Ride-pooling combines trips along similar routes into the same vehicle. Two vehicles (black lines) starting at the green and cyan origin, respectively, serve all requests.
    \textbf{c} Constraints modify the dynamics of the ride-pooling service. If only two customers can be transported by each vehicle at a time, the dark blue trip cannot be served as in panel b. Instead, the routes of the vehicles are modified and the customer is delayed. 
    }
    \label{fig:FIG1_ride_sharing_sketch}
\end{figure}

By combining trips of passengers along the same direction, ride-pooling reduces the required number of vehicles and the total distance driven. Similar to standard ride-hailing, on-demand ride-pooling services typically act as door-to-door transport for passengers, matching similar passenger requests to each other or to vehicles already on route, ideally without any detour for the passengers (Fig.~\ref{fig:FIG1_ride_sharing_sketch}a,b). In contrast to ride-hailing services, however, the assignment of passenger requests to ride-pooling vehicles is much more complex \cite{santi2014_shareabilityNetworks, alonso2017demand} due to the restrictions of the routes of the vehicles by already assigned passengers. The resulting complex collective dynamics of the ride-pooling fleet \cite{Molkenthin2019_TopologicalUniversality, lotze2022dynamic} and the intricate dependence of the service efficiency on the system parameters \cite{santi2014_shareabilityNetworks, tachet2017scaling, vazifeh2018_minimumFleetProblem} are far from fully understood. Previous studies have analyzed the potential to pair passenger requests as a graph covering problem \cite{santi2014_shareabilityNetworks} and demonstrated a universal scaling of the theoretical potential to combine rides with similar origin and destination across empirical demand patterns from different cities \cite{tachet2017scaling}. 
Recently, similar scaling laws have been demonstrated also in a simplified dynamical model of ride-pooling in the special case of unlimited passenger capacity \cite{Molkenthin2019_TopologicalUniversality}. However, similar to restrictions from already accepted requests, capacity limits of ride-pooling vehicles constrain the assignment of new requests to vehicles. A request that cannot be served by a vehicle due to capacity constraints must be picked up and delivered by another vehicle, potentially causing route changes and additional delays (see Fig.~\ref{fig:FIG1_ride_sharing_sketch}c). Thus, even this simple constraint on individual vehicles may strongly affect the collective dynamics of the ride-pooling fleet as a whole and thereby also change the dynamic scaling laws.  

Here, we analyze the collective dynamics of ride-pooling fleets under capacity constraints and identify the effective number of vehicles available to serve a request as the relevant scaling parameter to characterize their efficiency. With this effective available fleet size, we map the dynamics of capacity-constrained ride-pooling fleets to an unconstrained system, generalizing the scaling laws of ride-pooling efficiency. Moreover, we develop a queueing theory description of the ride-pooling dynamics in a minimal model system that enables an approximate analytical calculation of the efficiency and the relevant scaling parameters. Together with a self-consistent mean-field approximation in more complex settings, we demonstrate the possibility of using the scaling law to estimate required fleet sizes. Overall, our results suggest that universal scaling laws of ride-pooling efficiency may hold across a much broader range of settings and constraints and may thus enable the a-priori optimization of ride-pooling fleet size, capacity, and other system parameters in previously unserviced areas.

\section*{Results}
\subsection*{Collective dynamics of ride-pooling}
The dynamics of the ride-pooling fleet depend on a large number of system parameters. The topology of the underlying street network $\mathcal{G}$ and the demand distribution $\rho$ in space determine the average trip distance $\left<l\right>$ across all requests. The demand distribution in time, characterized by the average request rate $\lambda$, determines the number of requests. The number of vehicles $B$ and their properties, such as the typical velocity $v$ or passenger capacity $\theta$, as well as the dispatcher algorithm $\mathcal{A}$, assigning requests to vehicles, critically determine the resulting routes of the vehicles and thereby the service quality. 

We simulate the dynamics of the ride-pooling service in a simplified model. Customers request transport from one node of the underlying street network $\mathcal{G}$ to another node uniformly randomly following a Poisson process with rate $\lambda$. Each request is immediately assigned to a vehicle, adjusting its planned route, such that the request is delivered as fast as possible without delaying previous requests or exceeding the capacity constraints of the vehicles. Over time, vehicles drive along their planned routes, picking up and dropping off passengers, and the system settles into a steady operating state such that the average number $\left<C\right>$ of scheduled requests per vehicle (on board or scheduled to be picked up in the future) becomes constant if the system does not overload (Fig.~\ref{fig:FIG2_capacitylimit}a). We simulate these dynamics on various different network topologies, including simple network structures such as a minimal two-node graph or a complete graph, effectively one-dimensional topolgies in cycle graphs, as well as two-dimensional square lattices and geometric random networks. A more detailed description of the ride-pooling model and simulation parameters is provided in the Methods.

To compare the dynamics across different settings, we define the normalized load \cite{Molkenthin2019_TopologicalUniversality} 
\begin{equation}
    x = \frac{\lambda \left<l\right>}{vB} \,,
    \label{eq:dimensionlessLoad}
\end{equation}
describing the total average requested trip distance $\lambda\left<l\right>$ per time relative to the maximal distance $vB$ that all vehicles can drive. The load $x$ is a lower bound for the average occupancy of the ride-pooling vehicles. When $x > 1$, more distance is requested from the system than the vehicles can drive and ride-pooling is necessary to serve all requests. Stable operation of a ride-pooling service with maximum passenger capacity $\theta$ per vehicle is, in principle, possible for loads $x < \theta$. The service necessarily overloads for $x > \theta$ since each vehicle would need to transport more than $\theta$ customers on average to serve all requests. 

\subsection*{Capacity-unconstrained ride-pooling efficiency}

\begin{figure*}[h]
        \centering
        \includegraphics{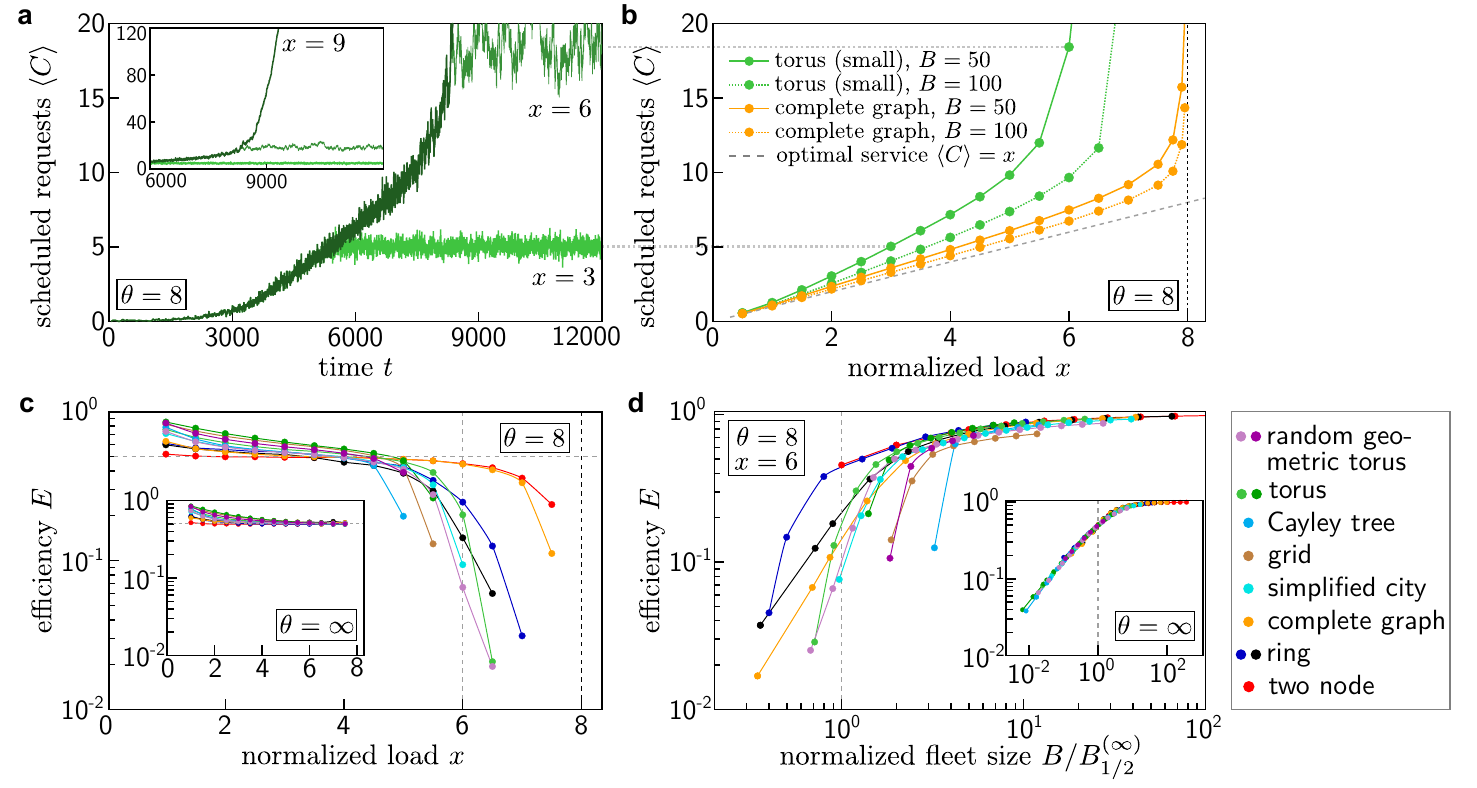}
        \caption{
            \textbf{Capacity constraints break the topological universality of ride-pooling efficiency.} 
            \textbf{a} The average number $\left<C\right>$ of scheduled customers per vehicle settles into a steady state for $x < \theta$ when the normalized load $x$  [Eq.~\eqref{eq:dimensionlessLoad}] is slowly increased. If the normalized load is larger than the capacity, $x > \theta$, the system overloads and the number of scheduled customers increases indefinitely (inset). 
            \textbf{b} For small loads, the average number of scheduled customers per vehicle increases approximately linearly with the normalized load $x$. The difference to the best possible scaling $\left<C\right> = x$ (dashed line) quantifies the efficiency of the service (see panels c and d). When the load $x$ approaches the capacity limit $\theta$, the number of scheduled customers diverges as the system overloads. 
            \textbf{c} Capacity-constrained systems behave qualitatively differently across network topologies when the load $x$ approaches the capacity limit $\theta = 8$ of the system. 
            (inset) Systems with unlimited vehicle capacity converge to the same efficiency $E$ for large loads $x$. Fleet sizes in both simulations are identical and chosen such that the efficiency $E$ of the capacity unconstrained systems (inset) converges to $E = 1/2$.
            \textbf{d} The efficiency curves $E_\mathcal{A}(\mathcal{T}, B, \theta , x)$ of the capacity-constrained systems reveal strong differences between the various network topologies (colors), especially in settings with small fleet sizes. Neither the normalized topological factor $B^{(\infty)}_\mathrm{1/2}(\mathcal{T})$ nor a load-dependent scaling factor $B_{1/2} \left(\mathcal{T},x\right)$ is sufficient to recover the topological universality observed for capacity-unconstrained systems [inset, Eq.~\eqref{eq:previousuniversality}]. Colors represent different underlying networks, see Methods for details on the settings and simulations. 
        }
        \label{fig:FIG2_capacitylimit}
\end{figure*}

The efficiency of a ride-pooling service can be consistently quantified across different settings based on the collective dynamics of the ride-pooling fleet \cite{Molkenthin2019_TopologicalUniversality}. If the capacity constraints of the system are sufficient to serve all requests, the system settles into a steady operating state with a constant number $\left<C\right>$ of scheduled requests per vehicle (Fig.~\ref{fig:FIG2_capacitylimit}a). The exact value of $\left<C\right>$ depends on the underlying network topology and system parameters (Fig.~\ref{fig:FIG2_capacitylimit}b). Under ideal conditions, requests are picked up immediately and delivered on the direct route to their destination. In this optimal service limit, each vehicle transports exactly $x$ passengers on average. The average number of scheduled requests per vehicle is equal to the average occupancy and equal to the normalized load $\left<C\right>_\mathrm{opt} = \left<O\right>_\mathrm{opt} = x$. The actual number of scheduled requests $\left<C\right>$ in a given system is typically larger since customers may have to wait for pickup or may be subject to detours in the pooled rides. The difference of the number of scheduled requests $\left<C\right>$ with respect to the optimal service limit thus quantifies the efficiency (Fig.~\ref{fig:FIG2_capacitylimit}c,d) of the ride-pooling system as \cite{Molkenthin2019_TopologicalUniversality}
\begin{equation}
    E = \frac{x}{\left<C\right>} \in [0,1] \,. \label{eq:eff}
\end{equation}

In general, fewer vehicles or a higher request rate, i.e.\ an increasing normalized load $x$, reduce the efficiency of a ride-pooling system as more requests have to be served with fewer vehicles in the same amount of time, resulting in longer waiting times and potential detours. However, a system with higher request rate $\lambda$ \emph{and} more vehicles $B$ (keeping the normalized load $x$ constant) operates closer to the perfect service limit. More vehicles increase the options for assigning requests while the increased request rate results in more similar requests that can be easily pooled, thus adding fewer constraints per request to the routing problem (Fig.~\ref{fig:FIG2_capacitylimit}b, \cite{Molkenthin2019_TopologicalUniversality, tachet2017scaling, santi2014_shareabilityNetworks}). 
Importantly, the system efficiency $E$ as defined above is directly related to the average service time $\left<\Delta t_s\right>$ from the perspective of customers. During the average service time $\left<\Delta t_s\right>$ of a single customer, a vehicle cycles on average exactly once through all its scheduled customers, i.e. dropping off all $\left<C\right>$ customers that were scheduled earlier. During this time, a total of $\lambda\,\left<\Delta t_s\right>$ requests are made to the system on average, of which a fraction $1/B$ is assigned to a specific vehicle. In the steady operating state, the average number of scheduled customers is thus given by 
\begin{equation}
    \left<C\right> = \frac{\lambda\,\left<\Delta t_s\right>}{B} \,.
\end{equation}
Using Eq.~\eqref{eq:dimensionlessLoad} and \eqref{eq:eff}, the efficiency \begin{equation}
    E = \frac{x}{\left<C\right>} = \frac{xB}{\lambda\,\left<\Delta t_s\right>} = \frac{\left<l\right>}{v} \,\frac{1}{\left<\Delta t_s\right>} \label{eq:load_customer}
\end{equation}
thus also quantifies the service efficiency from the customer perspective \cite{Molkenthin2019_TopologicalUniversality}.

The resulting efficiency $E_\mathcal{A}(\mathcal{T}, B, x, \theta)$ of a ride-pooling system with dispatcher $\mathcal{A}$ is a function of an effective topology $\mathcal{T} = (\mathcal{G}, \rho)$ that combines the street network topology with the spatial demand distribution, the fleet size $B$, the normalized load $x$, and the capacity $\theta$ of the vehicles. For ride-pooling systems with unlimited capacity $\theta = \infty$, this efficiency follows a universal scaling function $f_\mathcal{A}$,
\begin{equation}
    E_\mathcal{A}(\mathcal{T}, B, x, \infty) = f_\mathcal{A} \left(\frac{B}{B_{1/2} \left(\mathcal{T},x\right)} \right) \,,
 \label{eq:previousuniversality}
\end{equation}
with a single scaling parameter $B_{1/2}\left(\mathcal{T},x\right)$ summarizing the effect of the topology and the demand distribution \cite{Molkenthin2019_TopologicalUniversality}. For sufficiently large loads $x > 1$ in the ride-pooling regime, the scaling parameter $B_{1/2} \left(\mathcal{T},x\right)$ becomes approximately constant and we replace it with a single value $B^{(\infty)}_{1/2}\left(\mathcal{T}\right)$ for each effective topology $\mathcal{T}$ (Fig.~\ref{fig:FIG2_capacitylimit}d inset). 

However, systems that behave similarly without a capacity limit, exhibit stark differences in their efficiencies after introducing capacity constraints (Fig.~\ref{fig:FIG2_capacitylimit}c,d). The capacity constraints seem to break the universality, especially as the system load approaches the capacity limit, $x \rightarrow \theta$ (Fig.~\ref{fig:FIG2_capacitylimit}c). In contrast to the capacity unconstrained systems (Fig.~\ref{fig:FIG2_capacitylimit}d inset, Eq.~\eqref{eq:previousuniversality} \cite{Molkenthin2019_TopologicalUniversality}), the resulting efficiency curves for the capacity-constrained systems do not collapse (Fig.~\ref{fig:FIG2_capacitylimit}d). For fixed values of $x$ and $\theta$ we find that the scaling is qualitatively different across topologies. 

\subsection*{Capacity-constrained ride-pooling efficiency}
Can we recover the topological universality under capacity constraints and, if so, which are the relevant scaling parameters?

To understand the effect of the capacity constraints on the ride-pooling efficiency we examine their impact on the vehicle dynamics. The pick up and delivery dynamics along a planned route of a vehicle remain unchanged for capacity-constrained systems as the route of a vehicle is planned with respect to its capacity (i.e.\ all planned pick-ups are always possible). The capacity constraints thus only affect the routes and the fleet dynamics by modifying the assignment of requests. 

\begin{figure}[h!]
        \centering
	    \includegraphics{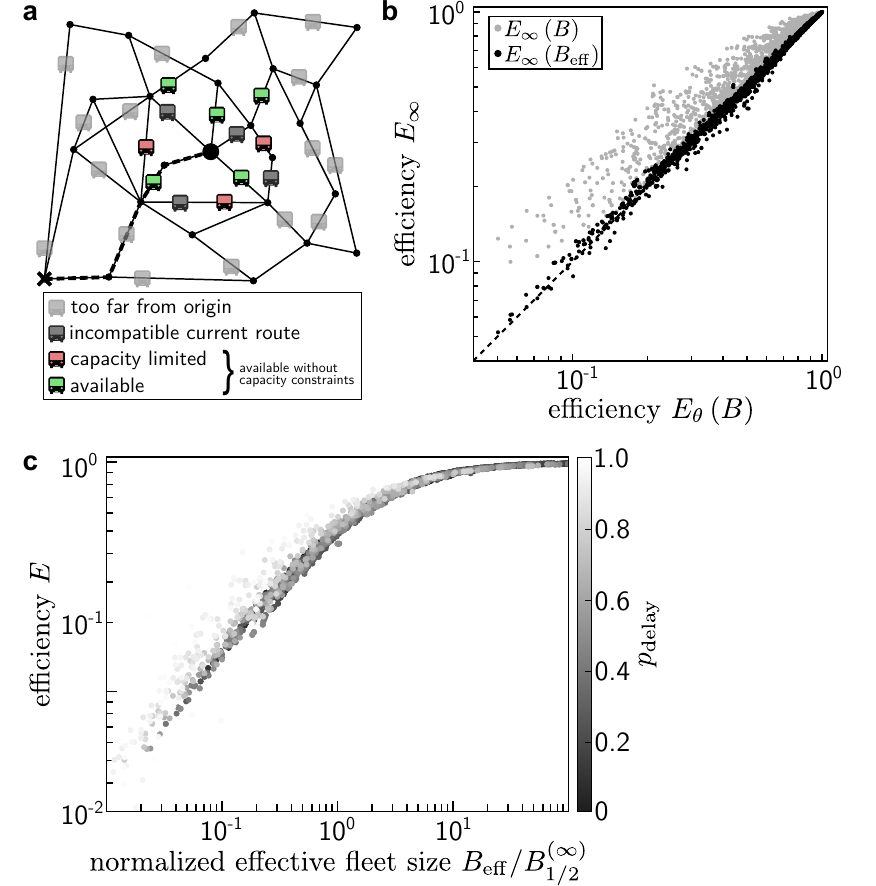}
        \caption{
            \textbf{Effective fleet sizes capture the impact of capacity constraints.} 
            \textbf{a} When a new request (black circle, center) arrives, it must be assigned to one of the ride-pooling vehicles in the system. The number of feasible vehicles to serve the request is limited due to the large distance to the origin of many vehicles (light gray) or incompatible planned routes of close-by vehicles (dark gray). In a system without capacity constraints, the request would be assigned to the best of the remaining vehicles. However, a fraction $p_\mathrm{delay}$ of these vehicles cannot serve the request due to the capacity constraints (light red). This argument suggests that the ride-pooling dynamics of a capacity-constrained system is similar to the dynamics of an unconstrained system with a reduced effective fleet size $B_\mathrm{eff} = (1-p_\mathrm{delay}) \, B$, Eq.~\eqref{eq:effective_fleet_size}. 
            \textbf{b} The efficiency $E_\theta(B)$ of capacity-constrained systems is approximately equal to the efficiency of unconstrained systems $E_\infty(B_\mathrm{eff})$ with the reduced effective fleet size $B_\mathrm{eff}$ (black dots). Comparing both systems with the same fleet size, the efficiencies differ significantly (light gray). The figure shows results for more than 3000 distinct settings $(\mathcal{T}, B, x, \theta)$ where $p_\mathrm{delay} \leq 0.8$. 
            \textbf{c} With the normalized effective fleet size as the scaling parameter, the efficiency of capacity-constrained ride-pooling services collapses to the same universal efficiency function as the unconstrained system across different topologies, capacity constraints, and system loads $x$. Deviations occur when most vehicles are fully occupied, $p_\mathrm{delay} \approx 1$ (light dots, see main text). See Methods for details on the settings and simulations.
        }
        \label{fig:FIG3_effective_fleet_size}
\end{figure}

Consider a system with a large fleet size and high efficiency. When a new request arrives, only vehicles that could serve the request with almost no delay are relevant options for the assignment (Fig.~\ref{fig:FIG3_effective_fleet_size}a). In both the capacity-constrained and unconstrained system, this excludes vehicles far away from the origin of the request. Similarly, vehicles close to the origin whose currently planned route is incompatible with the request are excluded since assigning the request to them would result in unfeasibly long waiting times or detours. Compared to the unconstrained system, capacity constraints further limit the pool of feasible options by excluding vehicles that would exceed their capacity constraints during the trip, thus resulting in longer delays. The dynamic routing decision effectively becomes identical to that of an unconstrained system without those unfeasible vehicles.

Assuming a homogeneous distribution of the unavailable, fully occupied vehicles among the pool of vehicles offering the most efficient trips, this argument suggests that the capacity-constrained system behaves similarly to a capacity unconstrained system with a reduced effective fleet size 
\begin{equation}
    B_\mathrm{eff}(\mathcal{T}, B, x, \theta) = \left[1-p_\mathrm{delay}(\mathcal{T}, B, x, \theta)\right]\,B \,. \label{eq:effective_fleet_size}
\end{equation}
This effective available fleet size characterizes the change in collective dynamics of the ride-pooling service due to capacity constraints. Consequently, the efficiency $E_\theta(B)$ of the capacity-constrained system is similar to the efficiency $E_\infty(B_\mathrm{eff})$ of an unconstrained system with the reduced fleet size $B_\mathrm{eff}$ (Fig.~\ref{fig:FIG3_effective_fleet_size}b). 
To quantify the fraction $p_\mathrm{delay}$ of unavailable vehicles, we measure the probability that the optimal assignment for a request is not possible due to the capacity constraints, i.e. the request is delayed compared to the capacity unconstrained system. 

This relation between capacity-constrained and -unconstrained ride-pooling dynamics suggests that the topological universality observed in unconstrained systems extends to capacity-constrained systems with the same scaling parameter $B_{1/2}$ and the effective fleet size $B_\mathrm{eff}$ (or equivalently the average fraction $p_\mathrm{delay}$ of unavailable vehicles) as a second scaling parameter. Figure~\ref{fig:FIG3_effective_fleet_size}c illustrates the collapse of the efficiency curves to a generalized universal scaling function 
\begin{equation}
    E_\mathcal{A}(\mathcal{T}, B, x, \theta) = f_{\mathcal{A}} \left(\frac{B_\mathrm{eff}(\mathcal{T}, B, x, \theta)}{B_{1/2} (\mathcal{T},x)} \right) \,
 \label{eq:newuniversality}
\end{equation}
of a single parameter with $B_\mathrm{eff} = (1-p_\mathrm{delay})\,B$, recovering the scaling of the unlimited capacity system with $p_\mathrm{delay} = 0$ (Fig.~\ref{fig:FIG3_effective_fleet_size}c). In contrast to the scaling parameter $B_{1/2}$ describing the topological universality, the effective fleet size $B_\mathrm{eff}$ depends on all system parameters, $(\mathcal{T}, B, x, \theta)$.

This scaling relation holds even for systems operating under high loads up to large values of $p_\mathrm{delay} \lesssim 0.8$. 
In systems operating very close to the capacity limit with $p_\mathrm{delay} \rightarrow 1$ and possibly $B_\mathrm{eff} < 1$, this mapping to a capacity unconstrained system begins to break down as also vehicles far away from the origin or with large detours become relevant for the assignment. These deviations are more likely for systems with strongly limited vehicle capacity or with very few vehicles. 

\subsection*{Mean-field queueing theory predictions}
Analytical calculations in a minimal two-node model confirm our results. With two nodes at a distance $\left<l\right>$, vehicles travel back and forth between the nodes without detours for customers. A vehicle arrives at a single node every $2\left<l\right>/(vB)$ time units on average. From the point of view of the node, all vehicles are identical since they always drop off all current customers when arriving and transport up to $\theta$ customers requesting a trip from that node. If vehicles are distributed equidistantly and never idle, the queueing dynamics at each node effectively follows a queue with Poisson distributed requests, a deterministic service interval $2\left<l\right>/(vB)$ with batch service for at most $\theta$ customers at the same time, and a single server \cite{bailey1954}. The average queue length $\left<q\right>$ of this system as well as the full queue length distribution can be computed analytically (\cite{bailey1954}, see Supplementary Material for detailed calculations). 

\begin{figure*}
        \centering
        \includegraphics{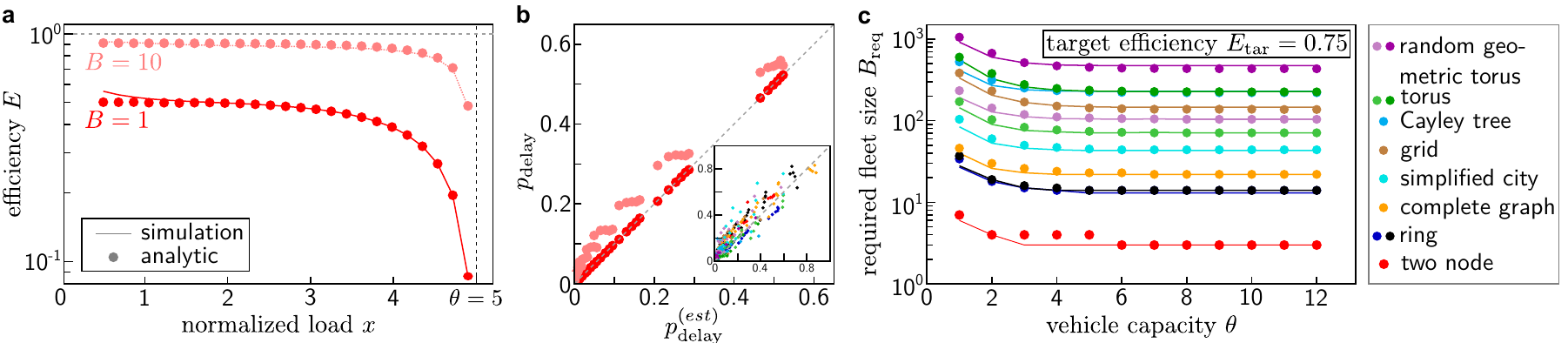}
        \caption{
            \textbf{Fleet size prediction for capacity-constrained ride-pooling services.} 
            \textbf{a} Queueing theory predictions (dots) of the ride-pooling efficiency in a minimal two-node setting. The predictions become exact for a single vehicle $B=1$ (dark red) at high load $x$ where the vehicle is never idle. Small deviations for larger fleet sizes ($B=10$, light red) reflect the non-equidistant inter-arrival time distribution of vehicles.
            \textbf{b} The same queueing theoretical description predicts the scaling parameter $p_\mathrm{delay}$ for various loads $x$ and capacity constraints $\theta$.
            (inset) A mean-field approach enables the estimation of $p_\mathrm{delay}$ in arbitrary networks for large numbers of vehicles (see Supplementary Material for details).
            \textbf{c} Prediction (dots) of the required fleet sizes to achieve a desired efficiency $E_\mathrm{tar} = 0.75$ for various network topologies and capacity constraints compared to direct numerical simulations (lines). These estimates rely only on the universal scaling function $f_{\mathcal{A}}$ and measurements of the scaling parameters $B_{1/2}(\mathcal{T},x)$ of the capacity unconstrained systems. 
            Colors represent different underlying networks, see Methods and Supplementary Material for details on the settings, simulations, and calculations.
        }
        \label{fig:FIG4_queueing_theory}
\end{figure*}

In the ride-pooling system, the average number $\left<C\right> = x + 2\left<q\right>/B$ of scheduled customers per vehicle consists of the number of customers currently transported per vehicle, $\left<O\right> = x$ since detours are impossible in this setting, and the queues at both nodes, $2\left<q\right>/B$. The efficiency becomes
\begin{equation}
    E = \frac{x}{\left<C\right>} = \frac{1}{1+2 \left<q\right> / (Bx)} \,,
\end{equation}
with a similar form as the universal scaling function predicted in \cite{Molkenthin2019_TopologicalUniversality}. This queueing theoretical prediction (Fig.~\ref{fig:FIG4_queueing_theory}a) becomes exact with $B=1$ vehicle for sufficiently large load $x$. For smaller loads, the vehicle becomes idle from time to time as fewer requests enter the system. For larger fleets, $B > 1$, fluctuations of the inter-arrival time lead to slight bunching of the vehicles and less efficient service.

The full queue length distribution from this model also provides direct access to the probability $p_\mathrm{delay}$ that a request is delayed due to the capacity constraints, i.e. when more than $\theta$ requests are waiting at a node when a vehicle arrives (Fig.~\ref{fig:FIG4_queueing_theory}b, see Supplementary Material for detailed calculations). As above, results are exact with $B=1$ vehicle. For larger fleets, fluctuations of the inter-arrival time and less efficient service result in more delayed requests and slightly larger values of $p_\mathrm{delay}$ than estimated.

Taking a mean-field approach and assuming that the queueing dynamics and occupancy statistics are identical at every node and vehicles arrive with a constant inter-arrival times in the limit of large fleets, the same approach also provides estimates $p^{(\mathrm{est})}_\mathrm{delay}$ for arbitrary networks (Fig.~\ref{fig:FIG4_queueing_theory}b inset). A detailed description of the estimation using a self-consistent solution of approximate queue length and occupancy distributions is given in the Supplementary Material. Differences between the estimated $p^{(\mathrm{est})}_\mathrm{delay}$ and the observed $p_\mathrm{delay}$ occur due to heterogeneities in the networks and the inter-arrival time of vehicles. As an alternative to an equidistant distribution of vehicles and a deterministic inter-arrival time, an exponential inter-arrival time distribution offers a good approximation for the dynamics in large and heterogeneous networks, reflecting the limit of many independent paths along which vehicles may arrive at a node (see Supplementary Material and Supplementary Figure S1).

Together with the scaling function $f_{\mathcal{A}}$, Eq.~\eqref{eq:previousuniversality} \cite{Molkenthin2019_TopologicalUniversality} and the topological factor $B_{1/2}$, this approximation enables us to a-priori estimate the required fleet size to achieve a desired efficiency in a given setting (Fig.~\ref{fig:FIG4_queueing_theory}c). Starting with some fleet size $B$, we estimate the delay probability $p_\mathrm{delay}$ and the effective fleet size $B_\mathrm{eff}$ using the mean-field calculations and compute the resulting efficiency $E$ from the universal scaling function. Comparing this estimate to a desired efficiency $E_\mathrm{tar}$, we obtain a new estimate for the required fleet size $B$ by assuming the same delay probability $p_\mathrm{delay}$. Iterating these estimations, the process converges to an estimate $B_\mathrm{req}$ of the required fleet size to achieve the desired efficiency in the given setting (see Supplementary Material for details). Note that, during this process, the load $x$ changes as the fleet size varies while the vehicle velocity, request rate, and request distribution remain constant. We thus make use of the full range of scaling parameters $B_{1/2}(\mathcal{T},x)$ of the capacity unconstrained systems to obtain more accurate results. For systems with a high density of requests, the topological factor $B_{1/2}(\mathcal{T},x)$ may be replaced by the single scaling factor in the limit of large loads $B^{(\infty)}_{1/2}(\mathcal{T})$, which can also be estimated without simulations in many simple networks by counting the number of distinct (shortest) paths \cite{Molkenthin2019_TopologicalUniversality}. 

The results of these estimations agree well with the required fleet sizes found from direct simulations in a wide range of network and capacity settings (Fig.~\ref{fig:FIG4_queueing_theory}c). Similar to the analytical calculations above, deviations become larger when $p_\mathrm{delay}$ is large (e.g. for low-capacity vehicles). However, this usually only occurs for undesirable settings with small target efficiencies or a large number of low-capacity vehicles. 

\section*{Discussion}
The collective dynamics of a ride-pooling fleet determines the potential and actual efficiency of the ride-pooling service \cite{tachet2017scaling, Molkenthin2019_TopologicalUniversality}. Instead of the specific request rate or the normalized loads, we have identified the effective number of available vehicles as the relevant scaling parameter to describe the dynamics of capacity-constrained ride-pooling fleets. This concept of an effective fleet size relates the efficiency of a capacity-constrained ride-pooling system to a system without capacity constraints and recovers the topological universality observed in systems with unlimited capacity \cite{Molkenthin2019_TopologicalUniversality}. The successful mapping between the collective dynamics of capacity-constrained and unconstrained systems suggests that a similar approach may be able to capture the impact of other constraints limiting the assignment of requests to vehicles, such as heterogeneous request sizes from individual travelers and groups or mixed request types for single (taxi cab) or shared rides. 

The universal scaling of the efficiency in systems without capacity constraints is robust across different demand distributions and network topologies (captured in the average trip length $\left<l\right>$ and the topological scaling factor $B_{1/2}$) as well as for different dispatcher algorithms in the high-efficiency limit \cite{Molkenthin2019_TopologicalUniversality}. Since our results are based on a direct mapping between capacity-constrained and -unconstrained systems, this robustness directly transfers as well. The mapping between the capacity-constrained and -unconstrained systems only breaks down for large $p_\mathrm{delay} \approx 1$ when the system is close to overloading, a state that is undesirable regardless of the setting due to long detours or waiting times. Since all arguments and in particular the definition of the ride-pooling efficiency rely on the equilibrium steady state of the ride-pooling dynamics, our results only capture expected dynamics over long times. Changes on timescales faster than the typical service time of a single customer, such as quickly changing or highly correlated demand distributions, strongly varying request rates $\lambda$, or quickly varying traffic conditions and vehicle velocities $v$, cannot be captured in this equilibrium description. Importantly, the scaling of the efficiency captures the dynamics both from the perspective of the provider in terms of the queueing theoretical throughput as well as from the perspective of the customers due to the direct relation to the average service time (see Eq.~\eqref{eq:load_customer}). A relevant additional perspective may be the extension of these scaling laws to the reliability of travel times and the distribution of delays beyond the mean-field description considered here. Similarly, while the dimensionless load quantifies when pooling rides becomes necessary, the sustainability of the service in terms of driven distance and emissions is not directly captured in the scaling laws. 

The analytic queueing theory model enables the application of this extended universality beyond numerical simulations. While the mean-field calculations for arbitrary networks cannot be expected to be highly accurate in real-life settings that are strongly heterogeneous, our results in principle enable a-priori estimates of required fleet sizes or efficiencies without the need for detailed simulations, complementing existing results \cite{santi2014_shareabilityNetworks, vazifeh2018_minimumFleetProblem, tachet2017scaling, Molkenthin2019_TopologicalUniversality} and providing a new tool to study the potential of ride-pooling in previously unserviced areas. 

\section*{Methods}

\subsection*{Ride-pooling simulations}
We simulate the dynamics of a ride-pooling service with $B$ vehicles traveling with constant velocity $v$. We set $v=1$ in all simulations without loss of generality, measuring time in appropriate units. For every vehicle, we store the planned routes as a list of scheduled pick-up and drop-off stops. Over time, vehicles drive along the shortest path between consecutive stops and pick up and drop off all scheduled customers. If a vehicle has no scheduled customers, it becomes idle and does not move until it is assigned a new customer. 

Customers place requests to travel from one node $i$ to another node $j \neq i$, distributed uniformly randomly and independently across all nodes in the network. Requests follow a Poisson process in time with an total rate $\lambda$ across the network.

Each time a new request is made, the dispatching algorithm iterates over all pick-up and drop-off insertions in the planned routes of all vehicles to find the offer that minimizes the arrival time of the request without delaying any previously scheduled customers. In case of multiple options, the secondary and tertiary objectives are the minimization of the time that the customer spends inside the vehicle and choosing the vehicle with the highest current occupancy, respectively. For transporters with limited capacities, only those offers are considered for which the occupancy does not exceed the capacity limit at any time during the trip. 

We simulate the dynamics in a variety of different settings described below. Each setting is described by a tuple of fixed parameters including the network topology $\mathcal{G}$, the fleet size $B$, the normalized load $x$ (or equivalently the request rate $\lambda$) and the capacity limit $\theta$ that applies to all vehicles.

In every simulation, we first distribute the (initially idle) vehicles uniformly randomly across all nodes of the network. We simulate $2000 \, B$ but at least $10^5$ requests to obtain an initial equilibrium state. Starting from this state, we enable the measurement of observables and again simulate in steps of $2000 \, B$ but at least $10^5$ requests. We stop the simulation when the average number of scheduled customers $\left<C\right>$ over the last $100$ time units deviates less than $10\%$ from the total average $\left<C\right>$ over the whole measurement period. Only for Fig.~2b in the main manuscript, we slowly increase the load by $\Delta x = 0.05$ and simulated for $1000$ or $1000\,x$ requests, whichever is larger ($1000\,x$ requests correspond to $1000\,\frac{x}{\lambda} = 1000\,\frac{\left<l\right>}{vB} = 50$ time units with a fleet size of $B=50$ vehicles and an average requested distance $\left<l\right> = 2.5$ on the small torus illustrated in the figure).

\subsection*{Model networks} 
We simulate the ride-pooling dynamics on different street networks $\mathcal{G}$. Nodes of the network correspond to possible pick-up and drop-off locations for customers and edges correspond to streets, with the edge length $l(i,j)$ between nodes $i$ and $j$ denoting the distance between adjacent nodes. 
\begin{itemize}
\item A \textit{minimal graph} consisting of $N=2$ nodes with $l(1,2) = l(2,1) = 1$. 
\item A small and a large \textit{ring} with $N=25$ and $N=100$ nodes, respectively, where neighboring nodes $i$ and $j$ have the distance $l(i,j) = 1$. 
\item A \textit{complete graph} with $N=5$, $l(i,j) = 1$ for all $i \neq j$. 
\item A non-periodic square lattice (\textit{grid}) with $N=100$ nodes and $l(i,j) = 1$ for every edge. 
\item A small and a large periodic square lattice (\textit{torus}) with $N=25$ and $N=100$ nodes, respectively, and $l(i,j) = 1$ for every edge. 
\item A \textit{simplified city} with $N=16$ nodes, which resembles a spider web. Four rays point outwards from an imaginary center. Four nodes are placed on each ray. On every ray, each node is connected to its neighboring node(s) on the same ray. Furthermore, on each two adjacent rays, the closest nodes to the center are connected to each other, as well as the third-closest nodes to the center. $l(i,j) = 1$ for any two connected nodes $i,j$. 
\item A \textit{Cayley tree} with $N=46$ nodes and $l(i,j) = 1$ for every edge.
\item A small and a large \textit{random geometric torus} with $N = 25$ and $N=100$ nodes, respectively. The networks are generated from the Delaunay triangulation of $N$ points distributed uniformly at random in the unit square with periodic boundary conditions. $l(i, j)$ is given by the Euclidean distance between the connected points $i$ and $j$ with respect to the periodic boundaries.
\end{itemize}

\subsection*{Measuring $\mathbf{p_\mathrm{\textbf{delay}}}$}
For each request, the dispatcher finds both the best offer $O_\theta$ respecting the capacity constraints and the best offer $O_\infty$ ignoring the capacity constraints. We define $p_\mathrm{delay}$ as the fraction of requests for which the two assignments $O_\theta$ and $O_\infty$ differ in terms of the assigned vehicle, the pick-up or the drop-off time. A difference in any of these parameters implies that the best offer in the unconstrained system has become unavailable due to capacity constraints. Note that the probability $p_\mathrm{delay}$ is a measure over requests for a single vehicle each time, not a direct measure for the fraction of unavailable, fully occupied vehicles.

\section*{Data availability}
Data and code underlying the results in the manuscript and the Supplementary Material is availble in the public Github repository 'PhysicsOfMobility/capacity\_constrained\_pooling', \cite{ZechGitHub2022} \url{https://doi.org/10.5281/zenodo.6624420}.


\bibliography{references}

\section*{Acknowledgements}
We thank Debsankha Manik and Philip Marszal for helpful discussion and suggestions. M.T. acknowledges support from the German Research Foundation (Deutsche Forschungsgemeinschaft, DFG) through the Center for Advancing Electronics Dresden (cfaed). M.S. and M.T. acknowledge support from the Volkswagen Foundation (VWF, Volkswagenstiftung) under Grant. No. 99 720.

\section*{Author contributions statement} 
M.S. and N.M. conceived the research, R.M.Z. performed the simulations and analytical calculations and created the figures supported by M.S., all authors analyzed the results and wrote the manuscript. 

\section*{Competing interest}
The authors declare no competing interests.

\clearpage

\renewcommand{\figurename}{Supplementary Figure}
\renewcommand{\thefigure}{S\arabic{figure}}
\renewcommand{\thetable}{S\arabic{table}}
\renewcommand{\theequation}{S\arabic{equation}}

{
    \centering
    \Large{\textbf{Supplementary Material}}\\
}

\thispagestyle{empty}

\section{Supplementary Note 1: Minimal model queueing theory}
Consider a minimal model of a ride-pooling service on a network with $N=2$ nodes $i \in \{1,2\}$ with distance $l_{1,2} = \left<l\right>$ with a single vehicle $B=1$ with capacity $\theta$ driving back and forth between the nodes with constant velocity $v$. Requests follow a Poisson process with rate $\lambda$ independently and uniformly randomly from one node of the network to the other, i.e. requests appear with rate $\lambda/2$ independently at each node. The travel time for customers is constant since detours are impossible in the minimal topology. Any reduction in efficiency is due to waiting times of customers until they are picked up by the vehicle.

Since all customers currently on the bus are dropped off when the bus arrives at a node, the bus always begins a trip with its full capacity $\theta$ available. The dynamics of both nodes are symmetric and it is sufficient to consider the queueing dynamics at a single node.

In the limit of high load ($x \gg 1$ or $x \rightarrow \theta$), when the bus is almost never idle since there are always pending requests to be served, the bus departs from node $1$ with up to $\theta$ of all $q_1(t_k)$ waiting passengers. It returns after a round-trip time at $T_{k+1} = t_k + \Delta t = t_k + 2\,\left<l\right>/v$. During this time, $z_1(t_k)$ new requests have arrived at node $1$ following a Poisson distribution with average $\lambda\,\left<l\right> / v = x$. The bus again picks up up to $\theta$ of the now waiting $q_1(t_{k+1})$ customers and repeats the cycle. The queueing dynamics at the node is described by a queue with Poisson arrivals with rate $\lambda/2$ with deterministic service interval $\Delta t = 2\,\left<l\right>/v$ and a single server $B=1$ with batch service with capacity $\theta$ \cite{bailey1954}.

\subsection{Average queue length and ride-pooling efficiency}
To compute the ride-pooling efficiency $E = x/\left<C\right>$, we need to compute the average number of scheduled customers $\left<C\right>$, consisting of the currently waiting customers at each node, $2\left<q\right>$, and the customers currently on board of the vehicle, $x$, where $\left<q\right>$ denotes the time-averaged queue length of the $M/D^\theta/1$-queue at a single node. 

Following the calculation of \cite{bailey1954}, let $\mathbf{p}$ denote the (infinite dimensional) vector of probabilities $p_q$ to observe a queue length $q$ just before the bus arrives. In equilibrium, $\mathbf{p}$ satisfies the fixed point equation
\begin{equation}
\mathbf{p} = P \mathbf{p} \,, \label{eq:queueing_theory_fixpoint}
\end{equation}
where $P$ is a matrix of transition probabilities with $P_{qq^\prime}$ denoting the probability to observe a queue length $q$ when the queue had length $q^\prime$ at the last service interval. The entries $P_{qq^\prime}$ are Poisson probabilities of the form
\begin{equation}
    P_{qq^\prime} = \begin{cases}
                        0 & \mathrm{if}   \quad q < q^\prime - \theta \\[1mm]
                        \frac{x^q\,e^{-x}}{q!} & \mathrm{if}   \quad q^\prime \le \theta \\[2mm]
                        \frac{x^{\left[q - \left(q^\prime-\theta\right)\right]}\,e^{-x}}{\left[q - \left(q^\prime-\theta\right)\right]!} & \mathrm{else}
                    \end{cases} \label{eq:minimal_queue_length_transitions_probailites}
\end{equation}
Further calculation \cite{bailey1954} yields the probability generating function $G(z) = \left<z^q\right>$ of $\mathbf{p}$,
\begin{equation}
G(z) = \frac{(\theta-x) (z-1) \prod_{i=k}^{\theta-1} (z - z_k)/(1-z_k)}{z^\theta e^{x(1-z)} - 1},
\label{eq:minimal_probability_generating_function}
\end{equation}
where 
\begin{equation}
z_k = - \frac{\theta}{x} \cdot W_0 \left( - \frac{x}{\theta} \cdot \mathrm{exp} \left( - \frac{x + 2 \pi k i}{\theta} \right) \right)
\end{equation}
are the $\theta - 1$ complex zeros of
\begin{equation}
z^\theta e^{x (1-z)} - 1 = 0
\end{equation}
within and on the unit circle and $W_0$ denotes the principal branch of the Lambert W function. 
From the probability generating function Eq.~\eqref{eq:minimal_probability_generating_function}, the average queue length $\bar{q}$ \textit{just before the bus arrives at the node} follows as 
\begin{equation}
\bar{q} = \frac{\theta - (\theta - x)^2}{2 (\theta - x)} + \sum_{k=1}^{\theta-1} \frac{1}{1 - z_k} \,. \label{eq:minimal_queue_length_before_arrival}
\end{equation}
Since on average $x$ customers arrive in one service interval, the average queue length $\underline{q}$ \textit{just after the bus has departed} from the node is $\underline{q} = \bar{q} - x$.
From these queue lengths and the Poisson arrival process, it follows that the time-average $\left<q\right>$ of the queue length is
\begin{equation}
\langle q \rangle = \frac{\bar{q} + \underline{q}}{2} = \bar{q} - \frac{x}{2} = \frac{\theta}{2} \left( \frac{1}{\theta - x} - 1 \right) + \sum_{k=1}^{\theta-1} \frac{1}{1 - z_k} \,.
\label{eq:minimal_time_average_queue_length}
\end{equation}
This expression captures the divergence of the queue length as the system overloads when $x \rightarrow \theta$.

With this expression for the average queue length we obtain the average number of scheduled cusomters as $\left<C\right> = x + 2\left<q\right>$ and find the expression for the efficiency of the service,
\begin{equation}
E = \frac{x}{\langle C \rangle} = \frac{1}{1+2 \langle q \rangle / x} \,,
\end{equation}
in the limit of sufficiently large $x$ when the vehicles are not idle.

The above calculations directly transfer to a system with a larger fleet $B>1$ under the assumption that the vehicles are equidistantly distributed. At constant $x$, the interval between two vehicles arriving decreases by a factor $B$ and the request rate increases by a factor $B$, resulting in the same average number $x$ of requests per service interval. Thus, since the number of queued customers does not change, the average number of scheduled customers \emph{per vehicle} becomes $\left<C\right> = x + 2\left<q\right>/B$. The efficiency consequently follows [Eq.~(6) in the main manuscript]
\begin{equation}
E = \frac{1}{1 + 2 \langle q \rangle/(xB)} \,.
\label{effminimalcap}
\end{equation}
The assumption of equidistant vehicles does not hold exactly in practice. If a single vehicle is delayed, a large number of customers making a request in the interval until the delayed vehicle arrives experience an increase of the waiting time. Fewer customers requesting a ride in the time interval until the next vehicle arrives experience a shorter waiting time. Overall, the average waiting time increases. Consequently, any deviation from an equidistant distribution of vehicles results in lower efficiency than predicted.

This calculation underlies the results presented in Fig.~4a in the main manuscript

\subsection{Estimation of $p_\mathrm{delay}$}
The queueing theory description also provides a way to compute the probability $p_\mathrm{delay}$ that a request is delayed due to the capacity constraints. We again consider only a single node due to the symmetry of the dynamics and denote the four relevant discrete random variables, measured when a vehicle arrives at the node, as follows
\begin{itemize}
    \item $Z$ denotes the number of newly scheduled customers since last service.
    \item $D$ denotes the number of customers (out of the $Z$ new ones) which cannot be served by the next bus arriving because the capacity constraint would be violated.
    \item $Q$ denotes the length of the queue just before the bus arrives at the stop.
    \item $Q'$ denotes the queue length just before the \textit{previous bus} arrived at the stop.
\end{itemize}
$p_\mathrm{delay}$ is then defined as the fraction of delayed requests,
\begin{equation}
p_\mathrm{delay} = \frac{E(D)}{E(Z)} \,, \label{eq:minimal_pfull_definition}
\end{equation}
where $E(\cdot)$ denotes the expectation value. 
The number of newly arriving customers $Z$ is a Poisson random variable with expected value $E(Z) = \lambda\,\left<l\right> / (vB) = x$, assuming an equidistant distribution of vehicles as above. 

In order to compute the expectation value $E(D)$ of the number of delayed customers, we compute the marginal probability mass function $P(D=d)$ as the sum over the joint distribution for all possible values of $Z$, $Q$ and $Q'$
\begin{eqnarray}
P(D=d) &=& \sum_{z=1}^{\infty} \, \sum_{q=1}^{\infty} \, \sum_{q'=0}^{\infty} P(D=d,Q=q,Z=z,Q^\prime=q^\prime) \label{eq:delayed_customers_probability}\\
&=& \sum_{z=1}^{\infty} \sum_{q=1}^{\infty} \sum_{q^\prime=0}^{\infty} P(D=d| Q=q, Z=z, Q^\prime=q^\prime) \, P(Q=q| Z=z, Q^\prime=q^\prime) \, P(Z=z| Q^\prime=q^\prime) \, P(Q^\prime=q^\prime) \,. \nonumber
\end{eqnarray}

The last probability $P(Q^\prime=q^\prime)$ is directly given by the equilibrium queue length distribution $p_{q^\prime}$ Eq.~\eqref{eq:queueing_theory_fixpoint}
\begin{equation}
P(Q^\prime=q^\prime) = p_{q^\prime} \,.
\end{equation}

The number of arriving customers $Z$ is independent of the current state of the queue such that 
\begin{equation}
P(Z=z| Q^\prime=q^\prime) = P(Z=z) = k_z \,.
\end{equation}

If the newly arriving customers $Z$ and the previous queue length $Q'$ are known, $Q$ follows deterministically. The previous vehicle picked up up to $\theta$ customers from the $Q^\prime$ waiting customers and $Z$ new customers arrived (compare Eq.~\eqref{eq:minimal_queue_length_transitions_probailites}. We thus have
\begin{equation}
    Q = \begin{cases}
            Z & \mathrm{if}   \quad Q^\prime \le \theta \\
            Q^\prime - \theta + Z & \mathrm{else}
        \end{cases}
\end{equation}
customers in the queue and the probability reduces to
\begin{equation}
P(Q=q| Z=z, Q^\prime=q^\prime) =
    \begin{cases}
      \delta_{q,z} & \quad \text{if} \quad q' \leq \theta \\
      \delta_{q,(z+(q'-\theta))}        & \quad \text{if} \quad q' > \theta
    \end{cases}
\end{equation}
with the Kronecker delta $\delta_{i,j} = 1$ if and only if $i=j$.

The number $D$ of delayed customers follows similarly in three cases:
\begin{itemize}
\item[i)] If $Q' \leq \theta$, then $Q = Z$. Then $D = \mathrm{max}\left[0,Z-\theta\right]$ customers are going to be delayed to a later vehicles.
\item[ii)] If $\theta < Q' < 2 \theta$, there are $Q'-\theta < \theta$ customers remaining in the queue after the previous vehicle leaves that will be picked up by the next vehicle. From the newly arrived $Z$ requests, $\theta - (Q'-\theta)$ will also be served, whereas $D = \mathrm{max}\left[0,Z - (\theta- (Q' - \theta))\right] = \mathrm{max}\left[0,Z + Q' - 2\theta\right]$ customers are delayed further.
\item[iii)] If $Q' \geq 2\theta$, only requests which where in the queue previously are served by the next vehicle and all the new requests are delayed, $D=Z$.
\end{itemize}
The relevant conditional probability for delaying customers follows as
\begin{equation}
P(D=d| Q=q, Z=z, Q'=q') = 
    \begin{cases}
      \delta_{d,\mathrm{max}\left[0,z-\theta\right]} & \quad \text{if} \quad q' \leq \theta \\
      \delta_{d,\mathrm{max}\left[0,z + q^\prime - 2\theta\right]} & \quad \text{if} \quad \theta < q' < 2 \theta \\
      \delta_{d,z}        & \quad \text{if} \quad q' \geq 2\theta
    \end{cases}
\end{equation}
With Eq.~\eqref{eq:delayed_customers_probability}, splitting the summation over $q'$ into the three distinct cases yields
\begin{eqnarray}
E(D) &=& \sum_{d=0}^{\infty} d\,P(D=d) \nonumber\\
&=& \sum_{d=1}^\infty \,  \sum_{z=1}^{\infty} \, \sum_{q=1}^{\infty} \left[ \, \sum_{q^\prime=0}^{\theta} d \, \delta_{d,\mathrm{max}\left[0,z-\theta\right]} \, \delta_{q,z} \, k_z \, p_{q^\prime} \right. \nonumber\\
&& \quad\quad\quad\quad\quad  + \sum_{q^\prime=\theta+1}^{2\theta-1} \, d \, \delta_{d,\mathrm{max}\left[0,z + q^\prime - 2\theta\right]} \, \delta_{q,(z+(q^\prime-\theta))} \, k_z \, p_{q^\prime} \nonumber\\
&& \quad\quad\quad\quad\quad  \left. + \sum_{q'=2\theta}^{\infty} \, d \, \delta_{d,z} \, \delta_{q,(z+(q'-\theta))} \, k_z \, p_{q^\prime} \; \right] \,.
\label{eq:expected_number_of_delayed_customers_a}
\end{eqnarray}
Eliminating all terms with $d=0$ by adjusting the $z$ bounds and evaluating the sum over $d$ yields
\begin{eqnarray}
E(D) &=& \sum_{z=\theta+1}^{\infty} \; \sum_{q=1}^{\infty} \, \sum_{q'=0}^{\theta} \, (z-\theta) \, \delta_{q,z} \, k_z \, p_{q'} \nonumber\\
&& \quad + \sum_{q'=\theta+1}^{2\theta-1} \, \sum_{z=2\theta-q'}^{\infty} \, \sum_{q=0}^{\infty} \, (z+q'-2\theta) \, \delta_{q,z} \, k_z \, p_{q'} \nonumber\\
&& \quad + \sum_{z=1}^{\infty} \, \sum_{q=1}^{\infty} \, \sum_{q'=2\theta}^{\infty} \, z \, \delta_{q,(z+(q'-\theta))} \, k_z \, p_{q'} \,. \label{eq:expected_number_of_delayed_customers_b}
\end{eqnarray}
For each constellation of $(z,q')$ there is exactly one $q$ within the summation bounds that satisfies $\delta_{q,\cdot} = 1$ in each term such that
\begin{equation}
E(D) = \sum_{z=\theta+1}^{\infty} \; \sum_{q'=0}^{\theta} \, (z-\theta) \, k_z \, p_{q'} \; + \; \sum_{q'=\theta+1}^{2\theta-1} \, \sum_{z=2\theta-q'}^{\infty} \, (z+q'-2\theta) \, k_z \, p_{q'} 
+ \sum_{z=1}^{\infty} \, \sum_{q'=2\theta}^{\infty} \, z \, k_z \, p_{q'} .
\end{equation}
Replacing the infinite sum in the last term using the normalization condition $\sum_{q'=0}^{\infty} p_{q'} = 1$ simplifies the expression
\begin{equation}
p_\mathrm{delay} = \sum_{z=\theta+1}^{\infty} \, (z-\theta) \, k_z \, \sum_{q'=0}^{\theta} \, p_{q'} \; + \; \sum_{q'=\theta+1}^{2\theta-1} \, \sum_{z=2\theta-q'}^{\infty} \, (z+q'-2\theta) \, k_z \, p_{q'} + \left( 1 - \sum_{q'=0}^{2\theta-1} \, p_{q'} \right) \,,
\end{equation}
such that only the first probabilities $p_{q'}$ for $q' \leq 2\theta-1$ are required to evaluate the expression. To evaluate this expression numerically, we cut off the summation over $z$ at $z_\mathrm{max} = 50$ (compared to typical values of $\theta$ and $x$ less than ten) because of the sharp decay of the Poisson probability $k_z$. 

This calculation underlies the analytical results presented in Fig.~4b in the main manuscript.

\newpage
\section{Supplementary Note 2: Mean field queueing theory for arbitrary networks}
The general idea of the queueing theoretical calculations above can be extended to arbitrary networks with a mean field approach. Assuming the queueing dynamics at all nodes and all vehicles are effectively identical, we can map the above calculation to arbitrary networks with effective parameters. The more heterogeneous the setting in terms of network topology or demand distribution, the larger the deviations from these mean field assumptions. 

There are three main differences to the minimal model calculations:
\begin{itemize}
    \item A single vehicle receives $\lambda/B = xv/\left<l\right>$ requests per time interval. With the \textit{average distance between nodes} $\left<e\right>$ (mean edge length), the vehicle receives on average $x\,\left<e\right>/\left<l\right>$ new requests between stops. For large fleet sizes and sufficiently low delay-probability $p_\mathrm{delay}$, only requests originating at the next stop of the vehicle will be assigned to it. Thus, the expected number $E(Z)$ of newly arrived customers at the next node on its route is
\begin{equation}
E(Z) = \sum_{z=0}^\infty k_z = x\,\frac{\left<e\right>}{\left<l\right>} =: x_\mathrm{eff} \le x \label{eq:arbitrary_network_effective_load}
\end{equation}
where $x_\mathrm{eff}$ denotes the effective load parameter for the mean field calculations.
\item Additionally, in large networks, the inter-arrival times between vehicles at a node are not identical. The inter-arrival time distribution is often more similar to an exponential distribution, reflecting a large number of (almost) independent shortest paths along which vehicles can arrive at a node. We include this variable inter-arrival time by modifying the probability $P(Z=z)$ with an integral over all possible inter-arrival times $\Delta t$
\begin{equation}
    P(Z=z) = k^\mathrm{exp}_z = \int_0^\infty e^{-\Delta t} \, \frac{e^{-x_\mathrm{eff}\Delta t} \, (x_\mathrm{eff}\,\Delta t)^z}{z!} \, \mathrm{d}\Delta t = \frac{x_\mathrm{eff}^z \, \Gamma (z+1)}{z! \, (1+x_\mathrm{eff})^{z+1}} \,. \label{eq:arbitrary_networks_exponential_arrival_distirbution}
\end{equation}
The following calculations are independent of the exact choice of the inter-arrival time distributions as it only enters via $P(Z=z)$.
\item Finally, not all passengers are dropped of at every stop such that vehicles do not always have its full capacity $\theta$ available for the new requests. We thus have to track the occupancy statistics of the vehicles in addition to the queue length statistics at the node. 
\end{itemize}

Let $O$ denote an additional random variable describing the occupancy of a vehicle at a node immediately after it has dropped of all passengers. The vehicle then has $\theta - O$ seats available for requests from that node. Let $\mathbf{p}$ with entries $p_q$ denote the probability to observe a queue length $q$ (as above) and $\mathbf{\pi}$ with entries $\pi_o$ denote the probability to observe an occupancy $o$. Similarly to Eq.~\eqref{eq:queueing_theory_fixpoint} above, we assume a steady state where both probability distributions fulfill the fixed point equations
\begin{eqnarray}
\mathbf{p} &=& P \, \mathbf{p} \nonumber\\
\mathbf{\pi} &=& \Pi \, \mathbf{\pi}\,. \label{eq:arbitrary_networks_equilibrium_condition}
\end{eqnarray}
Note that these equations are coupled since the occupancy depends on the number of queued customers and the number of queued (and delayed) customers depends on the occupancy. 

To compute the transition probabilities, we neglect correlations between observables that go beyond one service interval. The transition probabilities $P$ for the queue lengths follow similar to the minimal model. We define the relevant random variables as above:
\begin{itemize}
    \item $Z$ denotes the number of newly arrived customers since last service.
    \item $Q$ denotes the length of the queue just before the vehicle arrives at the stop.
    \item $Q^\prime$ denotes the queue length just before the \textit{previous vehicle} arrived at the stop.
    \item $O^\prime$ denotes the occupancy of the last vehicle that arrived at the node.
\end{itemize}
The transitions probability is given as the marginal probability
\begin{equation}
P_{q,q^\prime} = P(Q=q \, | \, Q^\prime=q^\prime) = \sum_{z=0}^\infty \sum_{o^\prime=0}^{\theta} \, P(Q=q \,| \, Z = z, O^\prime = o^\prime, Q^\prime = q^\prime) \, P(Z = z \,| \, O^\prime = o^\prime, Q^\prime = q^\prime) \, P(O^\prime = o^\prime \, | \, Q^\prime = q^\prime) \,.
\end{equation}
We readily insert
\begin{equation}
P(Z = z \, | \, O^\prime = o^\prime, Q^\prime = q^\prime) = P(Z=z) = k_z
\end{equation}
and
\begin{equation}
    P(O^\prime = o^\prime \, | \, Q^\prime = q^\prime) = \pi_{o^\prime} \,,
\end{equation}
where the latter equation implicitly assumes that there is no correlation between the occupancy of the previous vehicle and the queue length at that time.
With given $O^\prime$, $Q^\prime$ and $z$, the queue length $Q$ is deterministic as 
\begin{equation}
    Q = \begin{cases}
        Z & \mathrm{if}   \quad Q^\prime \le \theta - O^\prime \\
            Q^\prime - \left(\theta-O^\prime\right) + Z & \mathrm{else}
    \end{cases} \,,
\end{equation}
resulting in 
\begin{eqnarray}
P_{q,q^\prime} &=& \sum_{z=0}^\infty \sum_{o^\prime=0}^{\theta - q^\prime} \delta_{q,z} k_z\,\pi_{o^\prime} + \sum_{z=0}^\infty \sum_{o^\prime=\theta-q^\prime+1}^{\theta} \delta_{z,q-q^\prime+(\theta - o^\prime)} k_z\,\pi_{o^\prime} \nonumber\\
&=& k_q, \sum_{o^\prime=0}^{\theta - q^\prime} \pi_{o^\prime} + \sum_{o^\prime=\theta-q^\prime+1}^{\theta} k_{q-q^\prime+(\theta - o^\prime)}\,\pi_{o^\prime} \,. \label{eq:arbitrary_networks_transition_queue_length}
\end{eqnarray}

The transition probabilities $\Pi$ of the occupancy are more complex and require the estimation of how many customers leave the vehicle at the stop. Since we cannot track individual customers, we introduce two new random variables, only tracking one step explicitly and treating all customers who drive longer than one stop identically:
\begin{itemize}
    \item $L_1$ denotes the number of customers that were picked up at the last stop and are dropped off at the current stop.
    \item $L_\infty$ denotes the number of customers in the vehicle for more than one stop that are dropped off at the current stop.
\end{itemize}
We again write the transition probability as the marginal probability
\begin{eqnarray}
\Pi_{o,o^\prime} &=& P(O=o \,| \,O^\prime=o^\prime) \nonumber\\
&=& \sum_{q'=0}^\infty \, \sum_{l_1 = 0}^{\mathrm{min}\left[q', \theta-o'\right]} \, \sum_{l_\infty = 0}^{o'} P(O=o \,| \, L_1=l_1, L_\infty=l_\infty, Q^\prime=q^\prime, O^\prime=o^\prime) \, P(L_1=l_1 \, | \, L_\infty = l_\infty, Q^\prime=q^\prime, O^\prime=o^\prime) \nonumber\\ 
&& \quad\quad\quad\quad\quad\quad\quad \times\; P(L_\infty = l_\infty \, | \, Q^\prime=q^\prime, O^\prime = o^\prime) \, P(Q^\prime=q^\prime \, | \, O^\prime = o^\prime) \,. \label{eq:arbitrary_networks_occupancy_transition_marginal}
\end{eqnarray}
Similar to the above calculation, we take $P(Q^\prime=q^\prime \, | \, O^\prime = o^\prime) = p_{q^\prime}$. 

Since there is no difference between customers, both $L_1$ and $L_\infty$ follow Binomial distributions $B(l_1; \mathrm{min}\left[q', \theta-o'\right] ,p_1)$ and $B(l_\infty; o^\prime, p_\infty)$, respectively, where the second argument described the total number of customers that could be dropped off (i.e. that were picked up at the last node or that remained in the vehicle) and the third argument denotes topology-dependent drop-off probabilities measured from simulations. Alternatively, estimates for these probabilities could be obtained by counting neighboring nodes, assuming customers travel along shortest paths in the limit of large fleet size.

Given all other quantities, the occupancy follows deterministicly as
\begin{equation}
    O = O^\prime + \mathrm{min}\left[Q^\prime, \theta - O^\prime\right] - L_1 - L_\infty \,.
\end{equation}
Inserting these probabilities into Eq.~\eqref{eq:arbitrary_networks_occupancy_transition_marginal} and evaluating the $\delta$ operators results in the final expression
\begin{equation}
    \Pi_{o,o^\prime} = \sum_{q^\prime=0}^\infty \, \sum_{l_1 = l^\mathrm{min}_1}^{l^\mathrm{max}_1} \, B\left(l_c ; \mathrm{min} (\theta - o^\prime, \, q^\prime\right), p_1) \, B\left(o^\prime - o + \mathrm{min} (\theta - o^\prime, \, q') - l_c; o^\prime, p_\infty\right) \, p_{q^\prime} \, \label{eq:arbitrary_networks_transition_occupancy}
\end{equation}
with 
\begin{eqnarray}
l^\mathrm{min}_1 &=& \mathrm{max} (0, \, \mathrm{min} (\theta - o^\prime, \, q^\prime) - o) \nonumber\\
l^\mathrm{max}_1 &=& \mathrm{min} (\theta - o^\prime, \, q^\prime) - \mathrm{max} (0, \, o^\prime - o) \,.
\end{eqnarray}

We numerically compute the distributions $\mathbf{p}$ and $\mathbf{\pi}$ by iterating Eq.~\eqref{eq:arbitrary_networks_equilibrium_condition} with the transitions proabbilities Eq.~\eqref{eq:arbitrary_networks_transition_queue_length} and \eqref{eq:arbitrary_networks_transition_occupancy} for 100 times starting from a random initial distribution. We re-normalize the distributions in each step such that they describes a mean occupancy $x$, accurate in the limit of large fleet sizes and high efficiencies. To facilitate the numerical implementation, we cut off the queue length distribution at $q_\mathrm{max} = 50$. The distribution of the occupancy is naturally bounded by the capacity $\theta$.

\subsection{Estimation of $p_\mathrm{delay}$}

Following the same approach as above, we compute $p_\mathrm{delay}$ via 
\begin{equation}
p^\infty_\mathrm{delay} = \frac{E(D)}{E(Z)}
\end{equation}
with 
\begin{equation}
    E(Z) = x^\prime
\end{equation}
from Eq.~\eqref{eq:arbitrary_network_effective_load} and 
\begin{equation}
    E(D) = \sum_{d=0}^\infty \; d \, P(d) = \sum_{d,z,q,q'=0}^\infty \, \sum_{o, o'=0}^\theta \; d \, P(d \, | \, q, z, o, o', q') \, P(q \, | \, z, o, o', q') \, P(z \, | \, o, o', q') \, P(o \, | \, o', q') \,  P(o' \, | \, q') \, P(q') \,.
\end{equation}
Substituting all conditional probabilities
\begin{eqnarray}
P(d \, | \, q, z, o, o^\prime, q^\prime) &=& 
	\begin{cases}
		\delta_{d, \, z - (\theta - o)} & \mathrm{if}\quad q^\prime \leq \theta - o^\prime\\
		\delta_{d, \, z - (\theta - o - (q^\prime-(\theta - o^\prime)))} & \mathrm{if}\quad \theta - o' \leq q' \leq 2\theta - o^\prime - o\\
		\delta_{d, \, z} & \mathrm{if}\quad q' \geq 2\theta - o^\prime - o
	\end{cases} \nonumber\\[3mm]
P(q \, | \, z, o, o^\prime, q^\prime) &=& 
	\begin{cases}
		\delta_{q, \, z} & \mathrm{if}\quad q' \leq \theta - o'\\
		\delta_{q, \, z + (q^\prime - (\theta - o^\prime))} & \mathrm{if}\quad q^\prime > \theta - o'
	\end{cases} \nonumber\\[3mm]
P(z \, | \, o, o^\prime, q^\prime) &=& k_z \nonumber\\
P(o \, | \, o^\prime, q^\prime) &=& \pi_o \nonumber\\
P(o^\prime \, | \, q^\prime) &=& \pi_{o'} \nonumber\\
P(q^\prime) &=& p_{q^\prime} 
\end{eqnarray}
and evaluating the sum analogously to the calculation in the minimal model, we arrive at the estimate
\begin{eqnarray}
p^\mathrm{est}_\mathrm{delay} &=& \frac{1}{x'} \, \sum_{o, o^\prime=0}^\theta  \pi_o \, \, \pi_{o^\prime} \, \left[ \, \sum_{z=\theta-o+1}^\infty \; (z-(\theta - o)) \, k_z \, \sum_{q^\prime=0}^{\theta-o'} p_{q^\prime} \right. \nonumber\\
&& \quad\quad\quad\quad\quad\quad\quad + \sum_{q^\prime=\theta-o^\prime+1}^{2\theta-o-o^\prime-1} \;\;  \sum_{z=2\theta-q^\prime-o-o^\prime+1}^\infty \; (z+q^\prime-2\theta + o + o^\prime) \, k_z \, p_{q^\prime} \nonumber\\
&& \quad\quad\quad\quad\quad\quad\quad \left. + x_\mathrm{eff} \, \left( 1 - \sum_{q^\prime=0}^{2\theta - o^\prime - o - 1} \; p_{q^\prime} \right) \;\right] \,, \label{eq:arbitrary_network_pfull_estimate}
\end{eqnarray}
which we evaluate as the minimal model results numerically.

\begin{figure}[h]
    \centering
    \includegraphics{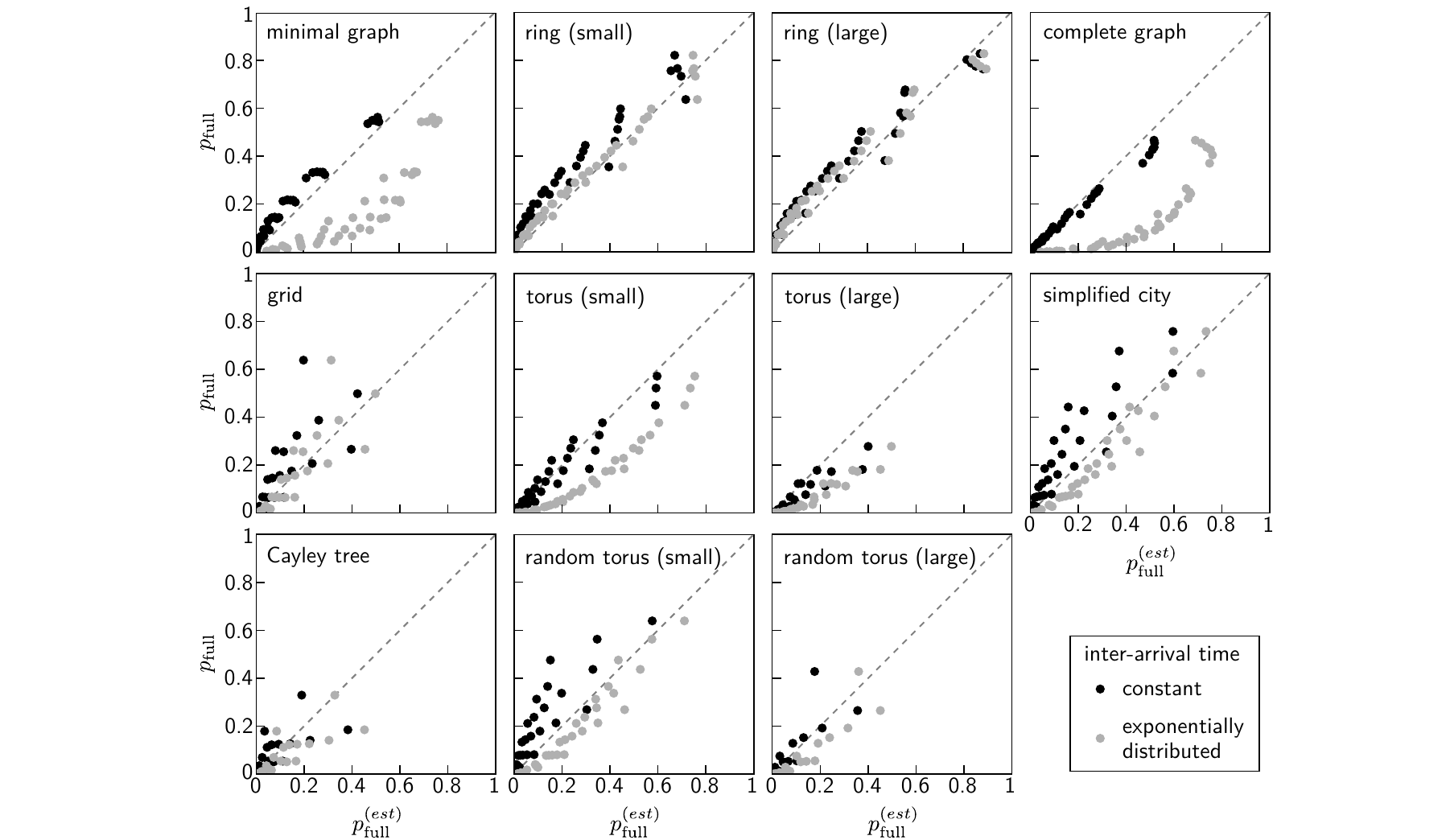}
    \caption{\textbf{Estimation of the delay probability $p_\mathrm{delay}$.} Black dots represent estimates of $p_\mathrm{delay}$ assuming equidistant arrivals of vehicles, gray dots represent the same estimate assuming an exponential inter-arrival time distribution. See Methods in the main manuscript for details on the settings and simulations.}
    \label{fig:FIGS1_p_full_estimation}
\end{figure}

This calculation with a deterministic inter-arrival time distribution (Poisson distributed new arrivals $Z$) underlies the results presented in the inset of Fig.~4b in the main manuscript. A comparison between estimations with equidistant arrivals and exponentially distributed inter-arrival times is shown in Fig.~\ref{fig:FIGS1_p_full_estimation}.

\subsection{Estimating $\mathbf{E_\infty(B_\text{eff})}$.} In order to compare a constrained system with its unconstrained equivalent, we calculate the effective fleet size $B_\mathrm{eff} = (1-p_\mathrm{delay})\,B$. To compute $E (\mathcal{G}, B_\mathrm{eff}, x, \theta = \infty)$ at potentially non-integer values $B_\mathrm{eff}$, we interpolate between efficiencies as follows.

For each $(\mathcal{G}, x)$, we select several fleet sizes $B'$ and find the corresponding efficiencies $E (\mathcal{G}, B', x, \theta = \infty)$ by simulation. In search of a regression function $E^\mathrm{\sim}_\infty (\mathcal{G}, \, \cdot \, , x, \theta=\infty)$ that fits this data, a multilayer perceptron with hidden layers of sizes $(5,10,5)$ and a $\mathrm{tanh}$ activation function has been trained with an $L2$ regularization using \texttt{scikit-learn}. The values on the vertical axis of Fig.~\ref{fig:FIG3_effective_fleet_size}b are the outcomes of inserting $B$ or $B_\mathrm{eff}$ into this function $E^\mathrm{\sim}_\infty$. While this approach is not strictly necessary to interpolate the efficiency function, it has the additional advantage of compensating statistical fluctuations from the measured efficiencies.

\subsection{Estimating $B_\mathrm{req}$.} To estimate the required fleet sizes, we fix the graph $\mathcal{G}$, the vehicle capacity $\theta$, the vehicle velocity $v$ and the request rate $\lambda$. We compute an estimate of the universal scaling function for various loads $x$ via regression of all model topologies using a neural network of four hidden layers of sizes $(80, 20, 10, 5)$ and a $\mathrm{tanh}$ activation function. We start with an estimate $B_\mathrm{req}$ and the resulting estimate for the delay probability $p_\mathrm{delay}$ to compute the efficiency via the regression of the universal scaling function. Since the load $x$ changes as we vary the fleet size, we interpolate linearly between values of the scaling parameter $B_{1/2}(\mathcal{T}, x)$ if required. We vary the fleet size $B_\mathrm{req}$ until the predicted efficiency is equal to the target efficiency $E_\mathrm{tar}$.

\end{document}